\documentclass[%
superscriptaddress,
showpacs,preprintnumbers,
nobibnotes,
 amsmath,amssymb,
 aps,
 prl,
 longbibliography,
 lengthcheck,%
]{revtex4-1}

\usepackage{graphicx}
\usepackage{dcolumn}
\usepackage{bm}
\usepackage{hyperref}

\begin{document}

\title{Symbolic Complexity for Nucleotide Sequences: A Sign of  the Genome Structure}

\author{R. Salgado-Garc\'{\i}a} 
\email{raulsg@uaem.mx}
\affiliation{Facultad de Ciencias, Universidad Aut\'onoma del Estado de Morelos. Avenida Universidad 1001, Colonia Chamilpa, 62209, Cuernavaca Morelos, Mexico.
} 

\author{E. Ugalde} 
\affiliation{Instituto de F\'{\i}sica, Universidad Aut\'onoma de San Luis Potos\'{\i}, Avenida Manuel Nava 6, Zona Universitaria, 
78290 San Luis Potos\'\i , Mexico.} 
\date{\today} 

\begin{abstract}
We introduce a method to estimate the complexity function of symbolic dynamical systems from a finite sequence of symbols. We test such complexity estimator on several symbolic dynamical systems whose complexity functions are known exactly.  We use this technique to estimate the complexity function for genomes of several organisms under the assumption that a genome is a sequence produced by a (unknown) dynamical system. We show that  the genome of several organisms share the property that their complexity functions behaves exponentially for words of small length $\ell$ ($0\leq \ell \leq 10$) and linearly for word lengths in the range $11 \leq \ell \leq 50$. It is also found that the species which are phylogenetically close each other have similar complexity functions calculated from a sample of their corresponding coding regions.

\end{abstract}


\maketitle

During the last decade there has been an intense debate about what does complexity mean for biological organisms and how it has evolved. Moreover, the problem of how to measure such a complexity at the level of nucleotide sequences, has became a challenge for geneticists~\cite{lynch2003origins,adami2002complexity,adami2000evolution}. Even having some well defined mathematical measures of complexity (most of them coming from the dynamical systems theory),  there are several problems in implementing such measures in real scenarios. The main difficulty lies on the fact that, due to the finiteness of the sample, the statistical errors are generally very large and the convergence in many cases cannot be reached (see Ref.~\cite{koslicki2011topological} and references therein). 

Here we will be concerned with the complexity function $C(\ell)$ (particularly for genomic sequences) defined as the number of sub-words of length $\ell$ (lets us call $\ell$-words hereafter) occurring in a given finite string. The importance of estimating such a quantity lies on the fact that it should give some information about the structure of the considered string, or, in other words, the mechanisms that \emph{produce} such a string.  The problem of determining the complexity function for finite sequences (and in particular of genomic sequences) has been previously considered by several authors~\cite{koslicki2011topological,colosimo2000special}.
It was found that the complexity function for a finite string has a profile which is independent on how the string was produced~\cite{koslicki2011topological,colosimo2000special}. For small values of $\ell$ (approximately $\ell \leq 10$ for nucleotide sequences) the complexity is an increasing function of $\ell$, after that, it becomes nearly constant on a large domain, and eventually becoming a decreasing function that reach zero at some finite $\ell$. This behavior is actually a finite size effect. Indeed, if we would like to compute the complexity function for the string, we would need a very large sample in order to obtain a good estimation. Assume, for sake of definiteness, that we are producing a random sequence, from a finite alphabet, as a fair Bernoulli trial (i.e., with the invariant measure of maximal entropy on the \emph{full shift}~\footnote{The \emph{full shift} is the set of all the infinite, or semi-infinite, sequences of symbols}). If the produced word $\mathbf{x}$ were of infinite length, then, all the words of all the lengths would \emph{typically} be present. Indeed, counting directly the number of different $\ell$-words  appearing in $\mathbf{x}$ we would obtain $\#\mathcal{A}^\ell$ \emph{almost always}, where $\#\mathcal{A}$ stands for the cardinality of the alphabet $\mathcal{A}$. However, if the produced sequence $\mathbf{g}$ have a finite length (which occurs when we stop the process at some finite time) then the number of $\ell$-words appearing in $\mathbf{g}$ should be regarded as a random variable which depends on the number of trails. 
Then, to compute the value of the complexity from a finite sequence we need to have a large enough sample in order to have an accurate estimation. For example, if $\ell = 20$ and the alphabet has four elements, then, as we know for random sequences, the complexity $C(20) = 4^{20} \approx 10^{12}$. This means that  for estimating this number, we would need a string with a size at least of $10^{12}$ symbols. This example makes clear that the difficulty we face when we try to estimate the complexity function is the size of the sample. Bellow we will show that, even with a small sample we can give accurate estimations for the symbolic complexity by using an appropriate estimator.

\begin{figure}[ht]
\begin{center}
\scalebox{0.3}{\includegraphics{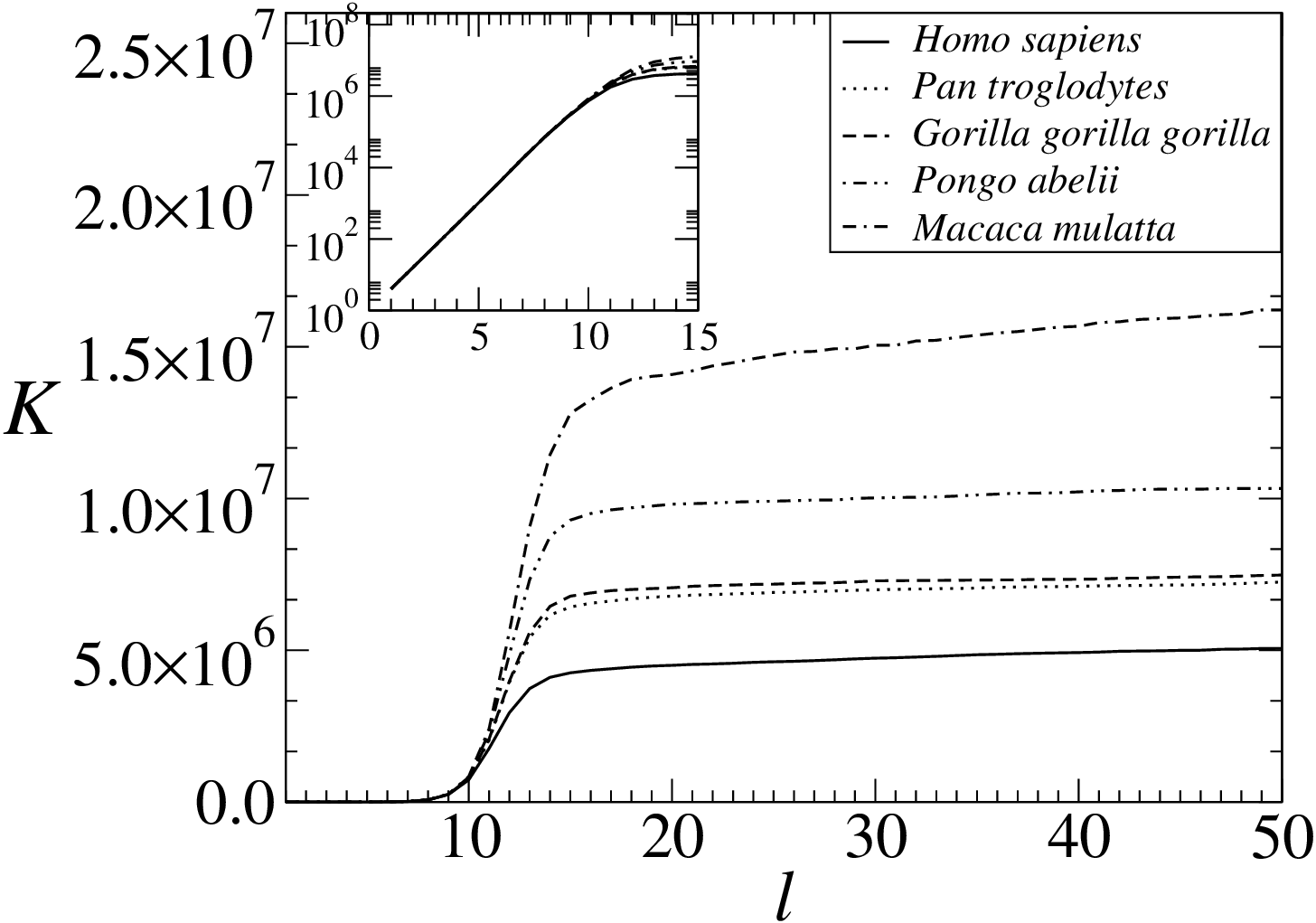}}
\end{center}
     \caption{ Complexity estimation for species belonging to \textit{Hominidae} family. The complexity was estimated from a nucleotide sequence of $6\times 10^6$ bp long taken from the first and from the second chromosomes (whenever necessary to complete the mentioned length) corresponding to coding regions. Then we taken a sample of $10^5$ $\ell$-words for $1\leq \ell \leq 50$. Here we can appreciate that the symbolic complexity corresponding to the human DNA is lower that rest of the \textit{Hominidae}.  Indeed, the order we observe according to the estimated complexity correlates qualitatively with the order in which they are found according to a phylogenetic distance estimated from other means (see for example~\cite{nei2000molecular}).  
                  }
\label{fig:Complexity_Hominidae}
\end{figure}

The point of view that we adopt here is to regard the complexity as an unknown property of a given stochastic system. Hence, this property has to be estimated from the realization of a random variable. The latter will be defined bellow and has a close relation with the number of different  $\ell$-words occurring in a sample of size $m$. In this way the proposed estimator lets us obtain accurate estimations for the  complexity finction of symbolic dynamical systems. 
We use this technique to give estimation of this symbolic complexity for coding DNA sequences. In Fig.~\ref{fig:Complexity_Hominidae}, we compare the symbolic complexity obtained from coding sequences of $6\times 10^6$ bp long (of the first chromosomes) of \emph{Homo sapiens}, \emph{Pan troglodytes}, \emph{Gorilla gorilla gorilla}, \emph{Pongo abelii} and \emph{Macaca mulatta} taken from the GenBank database~\cite{benson1997genbank}. From every sequence we taken a sample of $10^5$ words of lengths in the range $1-50$ bp. Then we calculated the corresponding values of $K$ for every $\ell$ which is our estimation of the symbolic complexity (see Eq.~\eqref{eq:estimator} bellow). In this figure we appreciate that the human coding sequences have the lowest complexity of all the species analyzed. From the same figure, we should also notice the progressive increasing of complexity as the species gets away from human, in the phylogenetic sense, according to the reported phylogenetic trees~\cite{nei2000molecular}. In such a figure we cal also appreciate a behavior which seems common to all organisms analyzed.  First,  we can observe that almost all the ``genomic words'' in the range $1-10$ are present in the (coding) nucleotide sequences analyzed. This is clear  from the exponential growth of words in this range  which fits to $C(\ell) \approx 3.94^\ell$ with a correlation coefficient $0.99$. Beyond the range $1-10$, our estimations let us conclude that the behavior of the complexity becomes linear. The latter suggest that the genomic sequences are highly ordered, or, in other words, the process by which this sequences are the result of a (quasi) deterministic one. In the literature it can be found that several symbolic dynamical systems having a linear complexity are actually the result of a substitutive  process,  like Thue-Morse, Toeplitz or Cantor sequences among others~\cite{allouche1994complexite,ferenczi1999complexity}.  Actually, the fact that the DNA could be the result of a random substitutive process has been suggested by several authors~\cite{li1991expansion,zaks2002multifractal,hsieh2003minimal,koroteev2011scale}.


Now lets us state the setting under which we give the estimator for the complexity.  Assume that a genome is produced by some stochastic process on a given symbolic dynamical system $(Y, \sigma)$. Here $Y \subset  \mathcal{A}^\mathbb{N}$ is a subset of semi-infinite symbolic sequences, made up from a finite alphabet $ \mathcal{A}$,  which is invariant under the shift mapping $\sigma$.  
Although the underlying dynamics producing the genome of a given individual is not known, we can assume that the set of allowed realizations of the genome $Y$ (the ``atractor'' of such a dynamics) can be characterized by a \emph{language}~\cite{lind1995introduction}. The language of a symbolic dynamical system is defined as the set of all the words of all sizes, appearing  in any point belonging to $Y$. If $\mathcal{A}_\ell$ the set of all the $\ell$-words appearing in any point $\mathbf{x}\in Y$, then the  \emph{language} of $Y$ is $\cup_{n\in \mathbb{N}} \mathcal{A}_n$. The symbolic complexity of $Y$ is then given by the cardinality of $\mathcal{A}_\ell$, i.e., $ C(\ell)  := \# \mathcal{A}_\ell$. Within this framework, a genome $\mathbf{g}$ of an individual can be considered as the observation of a point $\mathbf{x} \in Y$ with a finite precision.  Moreover, from such a point we can reconstruct the (truncated) orbit of $\mathbf{x}$ by applying successively the shift map to $\mathbf{g}$. If the sequence observed $\mathbf{g}$ is assumed to be typical with respect to some ergodic measure defined on the dynamical system (possibly an invariant measure of maximal entropy fully supported on $Y$), we can assume that the orbit generated by $\mathbf{g}$ explores the whole the attractor $Y$. Then, $\mathbf{g}$ must carry information about the structure of $Y$, and in particular of its symbolic complexity. As we saw above, the direct counting of words of a given length as  a measure of the complexity function requires a  large sample to have an accurate enough estimation. 

The problem we face can be stated as follows: given a sample of size $m$ of words of length $\ell$ we need to estimate the complexity $C(\ell)$ with the restriction $m < C(\ell)$ (and very often  $m \ll C(\ell)$ ). To this purpose, lets us assume that the words in the sample are randomly collected and that the realization of every word in the sample is independent from the rest. Let $Q$ be a random variable that counts the number of different words in the sample. It is clear that $1 \leq Q \leq m$. Under the assumption that all the words are equally probable to be realized in the sample, the probability function for $Q$ can be calculated exactly by elementary combinatorics,
\begin{equation}
f_Q(x) = \frac{\binom{m-1}{x-1}\binom{C}{x}}{\binom{C+m-1}{m-1}},
\label{eq:distributionQ}
\end{equation}
and the expected number of $Q$ can be calculated straightforwardly to give,
\begin{equation}
\mathbb{E}[Q] = \frac{Cm}{C + m -1}.
\label{eq:expectedQ}
\end{equation}
From the above we can see that, whenever the sample size $m$ is large enough compared to the complexity $C$ (the number of words of size $\ell$) the expected value of the random variable tends to the complexity $C$. The variance of $Q$ can also be calculated in a closed form, giving
\begin{equation}
\mbox{Var}[Q] = \frac{C m (C-1)(m-1) }{ (C+m-1)^2(C+ m -2)}. 
\label{eq:varQ}
\end{equation}
From this expression we should notice that the variance of $Q$ is small whenever $C\gg m$, and actually it goes as $\mbox{Var}[Q] \approx m^2/C$. This means that the deviations of $Q$ from its expected value are of the order of $m/\sqrt{C}$. In this regime, the expected value of $Q$ is approximately $m - \frac{m(m-1)}{C}$. We should notice from this asymptotic expressions that there is a regime in which the variance of $Q$ is small compared with the difference $\mathbb{E}[Q]-m$,  namely, when $ \sqrt{C}/m \ll 1$.  In this regime we have that almost any realization of $Q$ result in a value in which does not deviate significantly from  $\mathbb{E}[Q] \approx m - \frac{m(m-1)}{C} $ due to ``random fluctuations''. The latter is important since, as we can appreciate, it carry information about the complexity, which is in this case unknown. From this reasoning we propose the following estimator for the symbolic complexity $C$, 
\begin{equation}
K = \frac{m Q}{  m+1 - Q},
\label{eq:estimator}
\end{equation}
An few calculations shows that the expected value of $K$ is given by
\[
\mathbb{E}[K] = C + \frac{m^2-C^2}{m}\mathbb{P}(\{Q = m\}). 
\]
from which it is easy to see that proposed estimator $K$  is unbiased if $m>N$. 
We can see that, in the case in which the probability that all the words in the sample be different is small, any realization of $K$ is near $C$. 

Now, to implement this estimator to calculate the complexity we need to state how to meet the conditions imposed for the validity of the distribution given in Eq.~\eqref{eq:distributionQ}. We have to satisfy two main conditions: ($i$) that the words obtained in the sample be independent, and ($ii$) that the words of the same length have equal probability to occur. Lets us assume that a sequence $\mathbf{g}$ is a symbolic sequence of length $N$ obtained from some dynamical system. The orbit under the shift mapping generated by $\mathbf{g}$ can be written as $\mathcal{O}(\mathbf{g}) = \{ \mathbf{g}, \sigma(\mathbf{g}), \sigma^2(\mathbf{g}), \dots, \sigma^{N-1}(\mathbf{g})  \} $. A sample of words of length $\ell$ can be obtained from each point in the orbit by taking the first $\ell$ symbols. However, it is clear that the words obtained in this way are not independent. The latter is  due to the correlations between words generated by the overlapping when shifting to obtain the points in the orbit, and by the probability measure naturally present in the system which cause correlations even when two words sampled do not overlap. Thus, the sample should be taken from the orbit in such a way that the words are separated as most as possible along the orbit. Using this criterium we estimated the complexity for well known symbolic dynamical systems. First we produced long sequences of $6\times 10^6$ symbols from three different systems: the full shift (random sequences), the Fibonacci shift (sequences with the forbidden word $\mathrm{00}$), and the run-limited length shift (a \emph{sofic} shift, with a countable infinite set of forbidden words~\cite{lind1995introduction}). In every case the sequences were produced at random with the probability  measure of maximal entropy. Then we have taken a sample of $10^5$ words of lengths ranging from $1$ to $50$ for every sequence. Sampling in this way we have a separation of $10$ symbols between neighbor words of the maximal length analyzed $\ell = 50$.


\begin{figure}[h]
\begin{center}
\scalebox{0.3}{\includegraphics{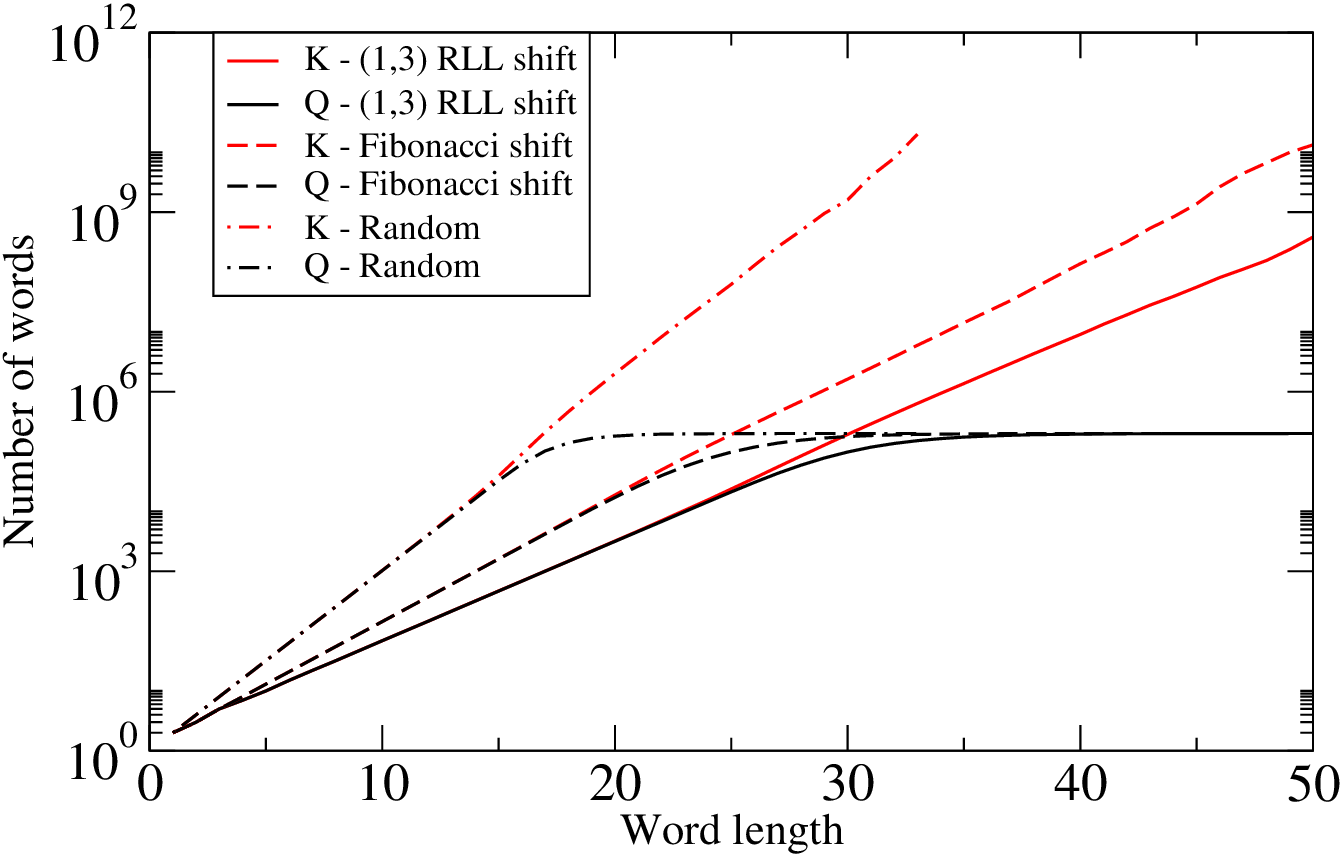}}
\end{center}
     \caption{ (Color online)
             Estimations for the complexity functions for three symbolic dynamical systems: the $(1,3)$-run-length limited (solid lines), the Fibonacci (dashed lines) and the full (dot-dashed lines) shifts. For each system we obtained a sequence of $6\times 10^6$ symbols by using the measure of maximal entropy. From such sequences we obtained samples of $10^5$ subwords of lengths ranging from $1$ to $50$. Then we obtained values for the random variables $K$ (red lines) and $Q$ (black lines) for every system.             
       From fits of the data shown for the random variable $K$ we estimated the respective complexities  of the form $C (\ell) \asymp  \exp(\hat h \ell)$. The estimated  values $\hat h$ are: $\hat h_{\mathrm{RLL} } =  0.384 \pm 0.0012$ for the $(1,3)$ run-length limited shift, $\hat h_{\mathrm{fib} } =  0.461 \pm 0.0014$ for the Fibonacci shift, and $\hat h_{\mathrm{rand} } =  0.721 \pm 0.0025$ for the full shift. The approximated  values obtained from analytical calculations are   $h_{\mathrm{RLL}} \approx 0.382$,  $h_{\mathrm{fib}} \approx 0.481$,  and   $h_{\mathrm{rand} } \approx 0.693$ respectively (see text).  
             }
\label{fig:Complexity_Shifts}
\end{figure}

In Fig.~\ref{fig:Complexity_Shifts} we show the values obtained for the random variables $Q$ and $K$ as functions of $\ell$ using the sample described above. From this figure we see that the values obtained for $Q$ as a function of $\ell$  exhibit a ``kink'', which has been previously observed in Refs~\cite{koslicki2011topological,colosimo2000special}. This behavior is consistent with the predicted by Eq.~\eqref{eq:expectedQ}, which can be calculated for these cases since we know the exact value of $C(\ell)$.  
Then, from the values of $Q$ we can obtain the values for $K$ which, as stated in Eq.~\eqref{eq:estimator}, gives an estimation for $C(\ell)$. It is known that $C(\ell)$ behaves exponentially in all the cases analyzed, i.e., $C (\ell) \asymp  \exp( h \ell)$, where $h$ is the topological entropy. It is known that the respective topological entropies are:  $h_{\mathrm{RLL}} = \ln(t^*)\approx 0.382$  for the (1,3)-run-length limited shift (where $t^*$ is the largest solution of $t^4-t^2 -t -1 = 0$), $h_{\mathrm{fib}} = \ln(\phi) \approx 0.481$ for the fibonacci shift (where $\phi $ is the golden ratio),  and   $h_{\mathrm{rand} } = \ln(2) \approx 0.693$ for the full shift~\cite{lind1995introduction}. From the curves for $K$ shown in the referred figure, we obtained the corresponding estimations for the topological entropies by means of the least squares method:  $\hat h_{\mathrm{RLL} } =  0.384 \pm 0.0012$, $\hat h_{\mathrm{fib} } =  0.461 \pm 0.0014$, and $\hat h_{\mathrm{rand} } =  0.721 \pm 0.0025$. From these results we observe that the better estimation made corresponds to the one for which the topological entropy is the lower. This is clear from Fig.~\ref{fig:Complexity_Shifts} since, due to the large number of words (especially in full shift) we have that the random variable $Q$ ``saturates'' rapidly, i.e., above some $\ell^*$ the expected value of $Q$ is differs in less than one, from the sample size $m$.


\begin{figure}[ht]
\begin{center}
\scalebox{0.30}{\includegraphics{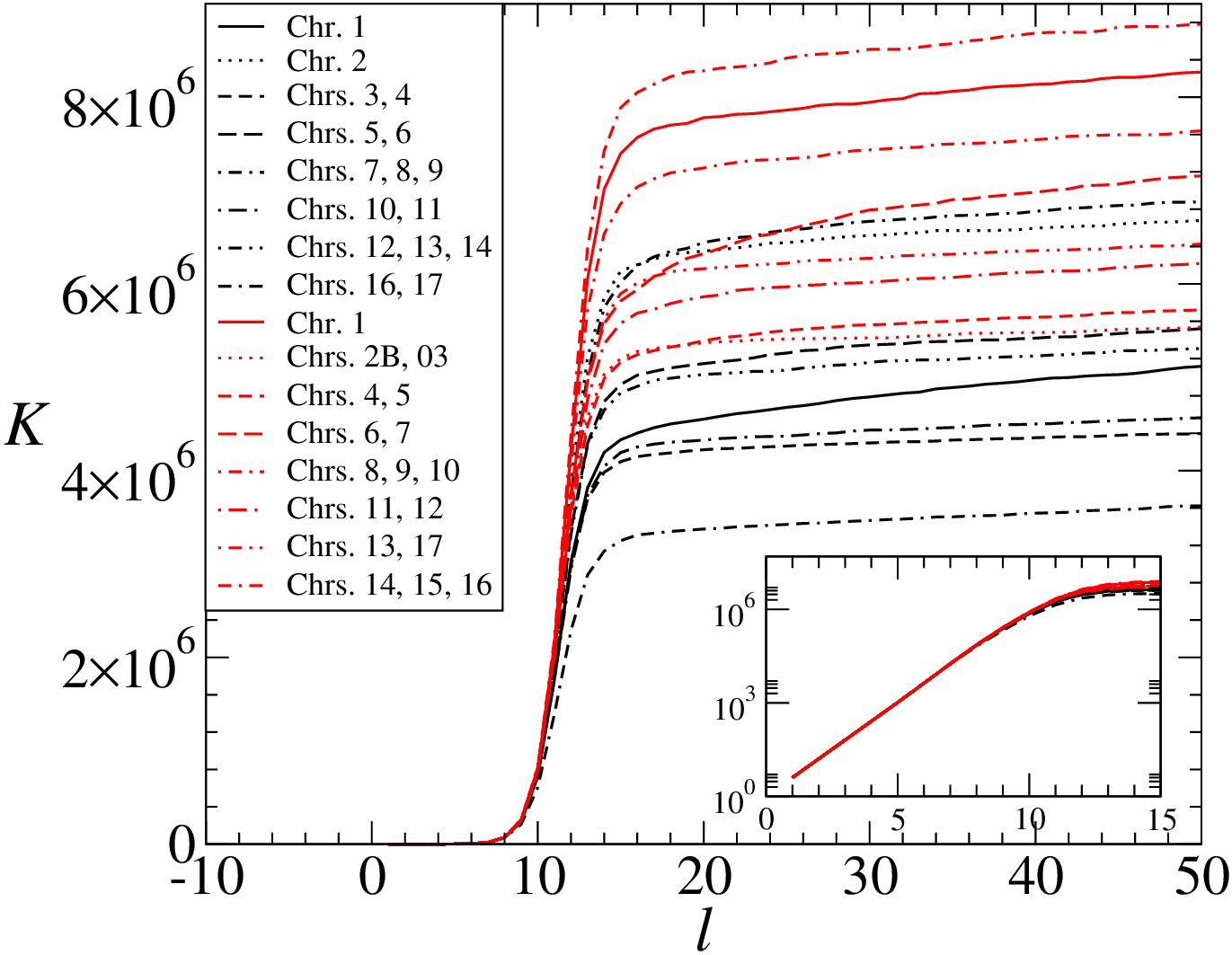}}
\end{center}
     \caption{ (Color online) The complexity function for the genome of the \textit{Homo sapiens} (black lines) and the \textit{Pan troglodytes} (red lines). Each curve corresponds to an estimation of the complexity function by means of the estimator given in Eq.~\eqref{eq:estimator}. For each curve we used one, two or more chromosomes in order to complete a sample string of $6\times10^6$ bp long.  From such a string we taken a sample of $10^5$ words of  $\ell$ bp for every $1\leq \ell \leq 50$.   
             }
\label{fig:complex_human_chimp}
\end{figure}

The reason for which we used coding DNA to estimate the complexity is due to the fact that the correlations on these kind of genomic sequences are practically absent in coding regions in the range 10-100 bp~\cite{buldyrev1995long,arneodo1998nucleotide,arneodo2011multi}.  This means that our hypothesis that the words in the sample be independent is at least fulfilled in the sense of correlations.  Even if we observe the behavior of the complexity in other regions of the genome (see Fig.~\ref{fig:complex_human_chimp} to appreciate the complexity functions for several chromosomes of \textit{Homo sapiens} and 	\emph{Pan troglodytes}), we found that the estimated complexity does not varies significantly from chromosome to chromosome. This also indicates that, at least in average, the coding regions seem to have a well defined complexity and therefore, a definite \emph{grammatical} structure in the sense of symbolic dynamics. 

In conclusion, we have proposed an estimator for the complexity function of symbolic dynamical systems. We tested such an estimator to calculate the complexity function of several symbolic dynamical systems whose complexity function is well known. Using this estimator we obtained the symbolic complexity for nucleotide sequences of coding regions of  genomes of four species belonging to the \emph{Hominidae} family. This study gave us information about the structure of the genome, which seems to be ubiquitous at least for all the species analyzed here. The main characteristic we found is that the complexity function behaves as mixture of exponential behavior (for words in the range $1$-$10$ bp) and an exponential one (for words in the range $11$-$50$ bp). This behavior is in some way consistent with 
several proposed evolution models that include a substitutive process since the linear complexity (which we observe for large genomic words) is a common characteristic of substitutive dynamical systems~\cite{allouche1994complexite,ferenczi1999complexity}
Moreover, the fact that the complexity does not varies significantly from chromosome to chromosome, suggest that there would exist a global architecture (a \emph{language} in the symbolic dynamics sense) for the coding region of the genome. It would be interesting to look for the (biological or dynamical) mechanisms responsible for the structure we found in the genomes of the \textit{Hominidae} family and if this structure is ubiquitous to the genomes of others organisms. We particularly found that the symbolic complexity correlates with the phylogenetic trees reported for these species. We believe that by analyzing the common features of the symbolic complexity several species could potentially be of help in the developing of whole-genome based phylogenetic reconstruction techniques. 

This work was supported by CONACyT through grant no. CB-2012-01-183358. 
R.S.-G. tanks F. V\'azquez for carefully reading the manuscript and for giving useful comments on this work.

\nocite{*}

\bibliography{StructGen.2012}

\end{document}